\documentclass[aps,preprint,superscriptaddress,longbibliography,prl]{revtex4-2}
\usepackage{graphicx, amsmath, verbatim, dsfont, amsfonts, color, amssymb, comment}
\usepackage[us]{datetime}
\usepackage{hyperref}

\graphicspath{{figures/}}

\newcommand*{\BS}{Bi$_2$Se$_3$}
\newcommand*{\BSES}{Bi$_2$Se$_3$/EuSe}
\newcommand*{\ESBS}{EuSe/Bi$_2$Se$_3$}

\begin{document}

\title{Discovery of a high-temperature antiferromagnetic state and transport signatures of exchange interactions in a \BSES\ heterostructure.}

\author{Ying Wang}
\affiliation{Department of Physics and Astronomy, Purdue University, West Lafayette, IN 47907 USA}
\author{Valeria Lauter}
\email{lauterv@ornl.gov}
\affiliation{Neutron Scattering Division, Neutron Sciences Directorate, Oak Ridge National Laboratory, Oak Ridge, Tennessee 37831, USA}
\author{Olga Maximova}
\affiliation{Department of Physics and Astronomy, Purdue University, West Lafayette, IN 47907 USA}
\author{Shiva~T. Konakanchi}
\affiliation{Department of Physics and Astronomy, Purdue University, West Lafayette, IN 47907 USA}
\author{Pramey Upadhyaya}
\affiliation{Elmore Family School of Electrical and Computer Engineering, Purdue University, West Lafayette, Indiana 47907, USA}
\author{Jong Keum}
\affiliation{Neutron Scattering Division, Neutron Sciences Directorate, Oak Ridge National Laboratory, Oak Ridge, Tennessee 37831, USA}
\affiliation{Center for Nanophase Materials Sciences, Physical Science Directorate, Oak Ridge National
Laboratory, Oak Ridge, Tennessee 37831, USA}
\author{Haile Ambaye}
\affiliation{Neutron Scattering Division, Neutron Sciences Directorate, Oak Ridge National Laboratory, Oak Ridge, Tennessee 37831, USA}
\author{Jiashu Wang}
\affiliation{Department of Physics and Astronomy, University of Notre Dame, Notre Dame, IN 46556, USA}
\author{Maksym Zhukovskyi}
\affiliation{Notre Dame Integrated Imaging Facility, University of Notre Dame, Notre Dame, IN 46556, USA}
\author{Tatyana~A. Orlova}
\affiliation{Notre Dame Integrated Imaging Facility, University of Notre Dame, Notre Dame, IN 46556, USA}
\author{Badih~A. Assaf}
\affiliation{Department of Physics and Astronomy, University of Notre Dame, Notre Dame, IN 46556, USA}
\author{Xinyu Liu}
\affiliation{Department of Physics and Astronomy, University of Notre Dame, Notre Dame, IN 46556, USA}
\author{Leonid~P.~Rokhinson}
\email{leonid@purdue.edu}
\affiliation{Department of Physics and Astronomy, Purdue University, West Lafayette, IN 47907 USA}
\affiliation{Elmore Family School of Electrical and Computer Engineering, Purdue University, West Lafayette,
Indiana 47907, USA}
\affiliation{Birck Nanotechnology Center, and Purdue
Quantum Science and Engineering Institute, Purdue
University, West Lafayette, Indiana 47907, USA}


\maketitle

\textbf{Spatial confinement of electronic topological surface states (TSS) in topological insulators poses a formidable challenge because TSS are protected by time-reversal symmetry. In previous works formation of a gap in the electronic spectrum of TSS has been successfully demonstrated in topological insulator/magnetic material heterostructures, where ferromagnetic exchange interactions locally lifts the time-reversal symmetry. Here we report an experimental evidence of exchange interaction between a topological insulator \BS\ and a magnetic insulator EuSe. Spin-polarized neutron reflectometry reveals a reduction of the in-plane magnetic susceptibility within a 2 nm interfacial layer of EuSe, and the combination of SQUID magnetometry and Hall measurements points to the formation of an antiferromagnetic layer with at least five-fold enhancement of N\'eel's temperature.
Abrupt resistance changes in high magnetic fields indicate interfacial exchange coupling that affects transport in a TSS. High temperature local control of TSS with zero net magnetization unlocks new opportunities for the design of electronic, spintronic and quantum computation devices, ranging from quantization of Hall conductance in zero fields to spatial localization of non-Abelian excitations in superconducting topological qubits.}

Topologically protected gapless surface states (TSS) are both a distinctive feature and a curse of topological insulators (TI)\cite{Hasan2011}: on the one hand, these states are robust against local disorder, on the other hand, they cannot be shaped by conventional lithographic techniques.  A gap in electronic spectrum of TSS can be induced by breaking time-reversal symmetry, e.g. by applying strong magnetic fields. Time-reversal symmetry can be spontaneously broken in a magnetic topological insulator: the quantum anomalous Hall effect (QAHE), a hallmark of broken time-reversal symmetry, has been observed both in intrinsic\cite{Deng2020} and conventional magnetically doped TIs\cite{Chang2013}. The disadvantages of bulk doping include the global gapping of TSS, uneven bandgap across the TSS due to inhomogeneous distribution of dopants, formation of impurity band and high density of lattice defects\cite{Liu2021}. The magnetic proximity effect\cite{Hao1990} is an attractive alternative to break time-reversal symmetry which alleviates the aforementioned drawbacks. There are a number of works where anomalous Hall effect has been reported in TI/magnetic insulator (TI/MI) heterostructures \cite{Wei2013,Yang2013,Alegria2014,Lang2014,Yang2014,Jiang2015,Katmis2016,Che2018,Fanchiang2018,Pereira2020,Wang2020,Allcca2022}, including TI/antiferromagnetic insulators\cite{Bhowmick2017,Pan2020}. In all these heterostructures magnetic exchange interaction induces out-of-plane ferromagnetic spin arrangement at the TI/MI interface; the exchange field breaks time-reversal symmetry and opens a gap in the TSS spectrum.

One of the most studied TI/MI system is EuS/\BS, where a combination of polarized neutron reflectometry (PNR) and superconducting quantum interference device (SQUID) magnetometry revealed interfacial magnetization that survives up to room temperature, a 20-fold enhancement of Curie temperature compared to the $T_c=17$ K in bulk EuS \cite{Katmis2016}. Subsequent studies by X-ray magnetic circular dichroism (XMCD) \cite{Figueroa2020} or low energy muon spin rotation experiments \cite{Krieger2019,Meyerheim2020} failed to detect an enhancement of interfacial $T_c$, most likely due to a low sensitivity or inadequate spatial resolution of these methods. Theoretically, the enhancement of the magnetic order was attributed to the presence of a strong spin-orbit coupling in TI\cite{Li2017}, however \textit{ab initio} calculations produced contradictory results.  Kim \textit{et al.} \cite{Kim2017} found that TSS in \BS\ can contribute to exchange coupling of Eu moments through RKKY mechanism that results in the enhancement of $T_c$ at the EuS/\BS\ interface. It was also argued that inter-diffusion of Eu across the interface  may induce magnetic moments on neighboring Se and Bi atoms an order of magnitude larger than the substrate-induced moments\cite{Eremeev2015}. On the contrary, a recent first-principles density functional study \cite{Tristant2021} did not find either additional induced magnetization at the interface or the magnetic proximity effect and a gap opening in electronic spectrum of TSS. Instead, their calculations revealed a downshift of energy states by 0.4 eV and relocation of TSS to the second quintuple layer (QL) due to partial charge transfer from Eu to Se and a dipole formation at the interface. Theoretical difficulties arise from the complex nature of the magnetic interactions in europium compounds\cite{Lee1984}. Experimentally, effective magnetic proximity coupling requires clean, sharp and controlled interface between a magnetic insulator and a TI due to the extreme short-range nature of the exchange interaction.

In this work we study interfacial exchange in \BSES\ heterostructures using a combination of PNR, SQUID magnetometry and electronic transport measurements. We present an experimental evidence of antiferromagnetic (AF) spin arrangement at the \BSES\ interface, where transport signatures of AF domain switching are observed at temperatures deep in the paramagnetic phase of the bulk EuSe. The use of AF materials to modify electronic transport through TSS states may be especially advantageous for spatial confinement of non-Abelian excitations, where zero net magnetization enables the use of aluminum, a low critical field superconductor with a large (1.3$\mu$m) coherence length.

\begin{figure}[t]
\includegraphics[width=0.98\textwidth]{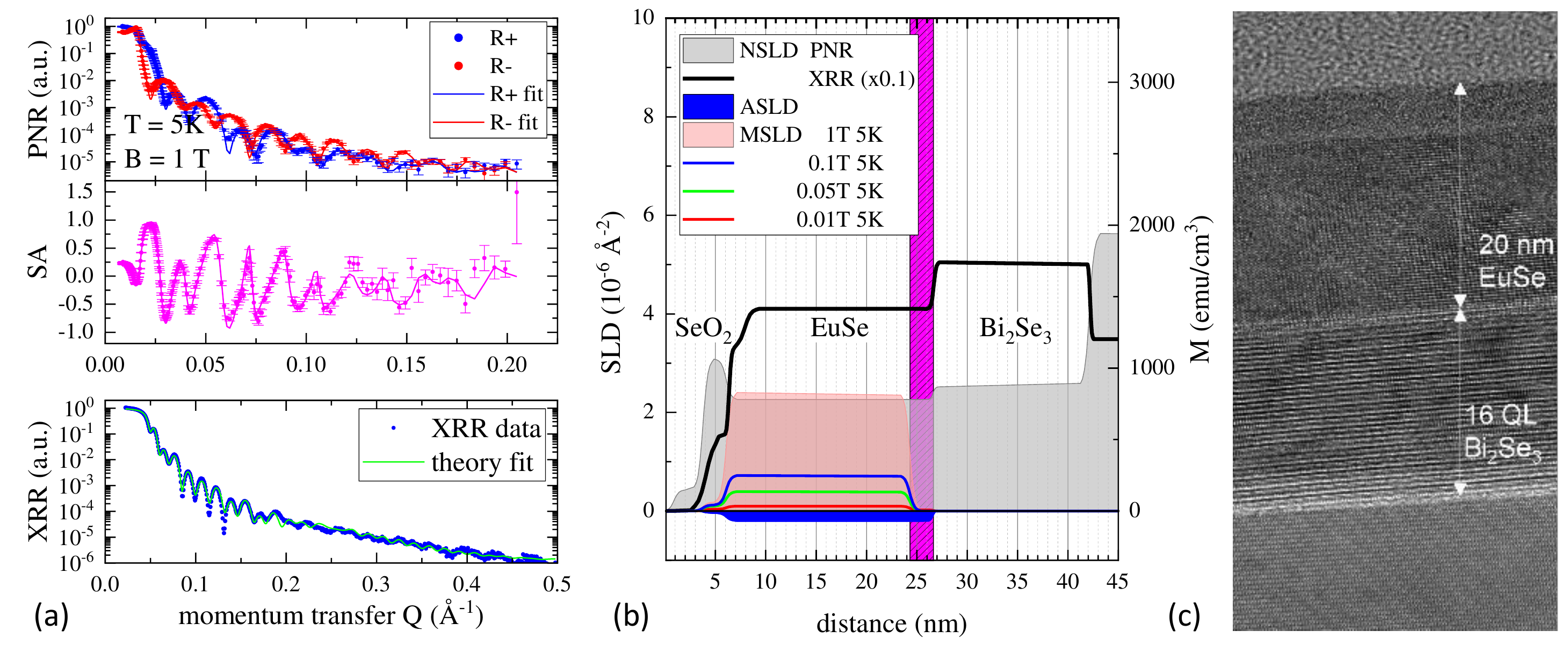}
\caption{\textbf{Polarized neutron reflectometry (PNR)}. (a) PNR data for spin-up ($R^+$) and spin-down ($R^-$) neutrons (top panel), spin-asymmetry ratio $SA=(R^+-R^-)/(R^++R^-)$ (middle panel), and X-ray reflectometry (XRR) data (bottom panel) are plotted as a function of the wave vector transfer $Q=4\pi\sin\theta/\lambda$ (where $\theta$ is the incident angle and $\lambda$ is the neutron wavelength). Solid lines are fits to the data, error bars represent corresponding standard deviations. (b) PNR nuclear (NSLD, grey), magnetic (MSLD) and absorption (ASLD, blue) scattering length density (SLD) profiles are plotted as a function of the distance from the sample surface. The red, green and magenta solid lines, as well as pink area show MSLD profiles at $H=0.001, 0.05,0.1$ and 1 T. The scale on the right corresponds to the magnetization profile. The interfacial layer of EuSe where magnetization is suppressed is highlight with a magenta color. (c) High resolution transmission electron microscopy (HRTEM) image of the sample.}
\label{f:f1}
\end{figure}

EuSe is an antiferromagnetic member of the Eu chalcogenides family with the N\'eel's temperature in the bulk $T_N=4.6$ K\cite{Mauger1986}. Near cancellation between nearest and next-nearest neighbor interactions results in a rich magnetic phase diagram which includes antiferromagnetic (AF), ferrimagnetic (FiM) and ferromagnetic (FM) phases. Eu spins lay within the (111) planes where they are coupled ferromagnetically, and AF, FiM and AF phases refer to the  mutual orientation of (111) planes. Thin EuSe films grown on \BS\ using molecular beam epitaxy prefer (001) crystal orientation and their magnetic phase diagram is found to be similar to the bulk material\cite{Wang2022a}. For this study, 20 nm of EuSe has been epitaxially grown on \BS. The materials form a sharp interface as evident from high resolution transmission electron microscopy images (HRTEM), Fig.~\ref{f:f1}c. While \BS\ grows as a monocrystal over a large surface, EuSe forms a domain structure with characteristic size 15-30 nm, see extended HRTEM images in the Supplementary Material.  From the temperature dependence of magnetic susceptibility at zero magnetic field $T_N\approx4.4$ K is obtained. At $H\sim 20$ mT AF-FiM transition is observed (the exact value depends on the direction of the magnetic field sweep), the film is in a FM phase above $0.2$ T, see Ref.~\cite{Wang2022a} and Supplementary Figure \ref{f:sf3}. This rich phase diagram allows investigation of the dependence of interfacial exchange interactions on the magnetic phase in the bulk of the EuSe film.

\begin{figure}[t]
\centering
\includegraphics[width=0.98\textwidth]{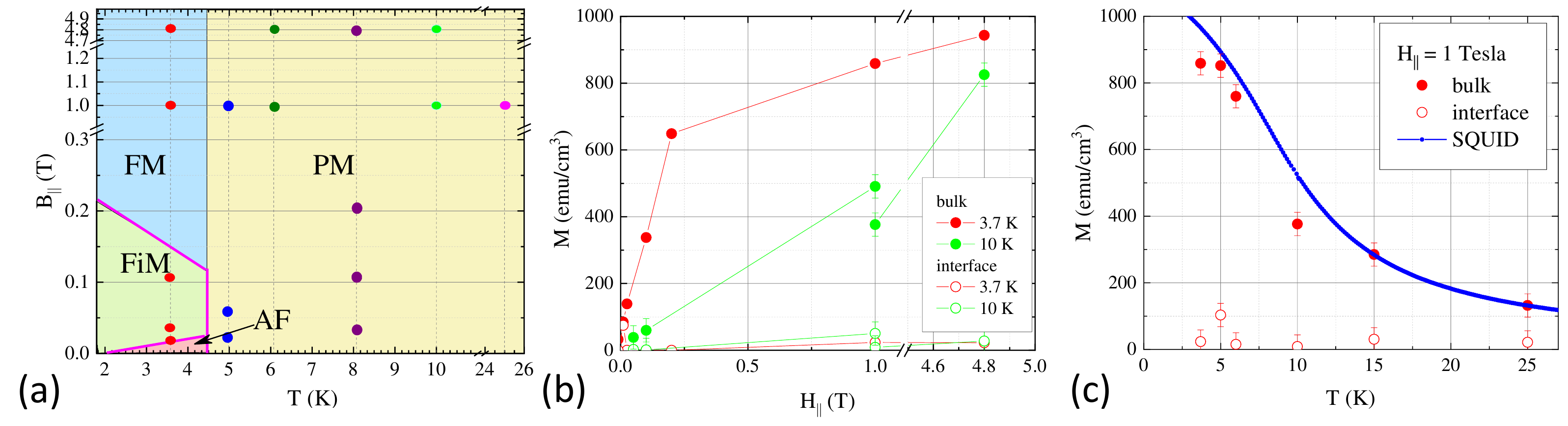}
\caption{\textbf{Temperature and field dependence of in-plane magnetization.} (a) Phase diagram of a EuSe thin film obtained using SQUID magnetometry (magenta lines) includes antiferromagnetic (AF), ferrimagnetic (FiM), ferromagnetic (FM) and paramagnetic (PM) phases. Color dots mark positions of PNR measurements. (b,c) Magnetic field and temperature dependence of bulk and interfacial in-plane magnetization extracted from PNR measurements. Overall magnetization measured using SQUID magnetometer is shown by a blue line. In (b) two sets of data at 10 K (0-1T and 1-4.8T) were taken 1 year apart and reflect small reduction of EuSe film magnetization with time. Note breaks in horizontal and vertical axes in plots (a) and (b).}
\label{f:f2}	
\end{figure}

We apply a method of polarized neutron reflectometry (PNR) as a unique tool that provides high resolution of chemical composition and depth profile of the in-plane magnetization vector and, thus, can probe structure and magnetization at a buried interface. A PNR  data of \BSES\ film measured in the presence of in-plane external magnetic field $H_\|=1$ T at $T=5$ K (paramagnetic phase of the bulk EuSe) is shown in Fig.~\ref{f:f1}a.  There is a visible difference in $R^+$ and $R^-$ reflectivities for neutrons with spins aligned parallel or antiparallel to the applied magnetic field ($R^+$ and $R^-$, respectively). In order to improve precision in the interface depth determination X-ray reflectivity (XRR) measurements were performed on the same sample at room temperature. The fit to the data was performed simultaneously for the PNR and XRR data. The depth profiles of the nuclear and magnetic scattering length densities (NSLD and MSLD), obtained from the fit to the data, correspond to the depth profile of the chemical and in-plane magnetization vector distributions, respectively.

The middle panel in Fig.~\ref{f:f1}a shows the spin asymmetry ratio $SA=(R^+ - R^-)/(R^+ + R^-)$ obtained from the experimental and fitted reflectivity profiles. The SA signal evidences the presence of a depth-dependent magnetic moment. Thus determined positions of \BSES\ and \BS/sapphire interfaces are consistent with the high resolution TEM (HRTEM) images of the sample's cross section. NSLD shows $\approx 10\%$ contrast between scattering on \BS\ and EuSe with a 1 nm transition region consistent with the PNR depth resolution. The accuracy of the \BS\ interface determination is further improved and confirmed through the analysis of the absorption scattering length density (ASLD), which comes solely from scattering on Eu atoms\cite{Korneev1992}. Both PNR and HRTEM data reveal a sharp \BSES\ interface, the interfacial roughness obtained from the fit of the PNR and XRR data is $0.6\pm 0.4$ nm, extending 1 monolayer (ML) into EuSe and less than 1 quintuple layer (QL) into \BS.

MSLD profile representing depth distribution of in-plane magnetization is plotted for $H_{\|}=0.01, 0.05, 0.1$ and 1 T in Fig.~\ref{f:f1}(b). Gradual increase of bulk magnetization with magnetic field is consistent with paramagnetic state of the EuSe film at $T=5$ K, magnetization saturates at $M_0\approx 7\mu_B$ per atom, where $\mu_B$ is the Bohr magneton, for $B>0.2$ T. At the same time there is a distinct suppression of magnetization in EuSe within $2\pm0.6$ nm from the interface. Note that this region is clearly within the EuSe layer according to both ASLD and NSLD data, as is highlight with a vertical shaded area in Fig.~\ref{f:f1}(b). To investigate further this feature, we performed PNR measurements in a broad range of magnetic fields and temperatures, see Fig.~\ref{f:f2} and Supplementary Material. From a detailed analysis of PNR data it follows that the suppression of the interfacial magnetization  is independent of the magnetic state of EuSe film: the data, collected at 3.7 K spans AF, FiM and FM phases of the bulk of the film, while the data collected at 10K corresponds to the PM phase. In Fig.~\ref{f:f2}b magnetization $M$ in the bulk and in the interfacial layer of the EuSe film is plotted as a function of in-plane magnetic field. At $T=3.7$ K the bulk $M$ reaches a FiM state ($\sim 1/3 M_0$) at $H_\|\approx0.1$ T and $>90\%$ of $M_0$ at $H_\|=4.8$ T. The interfacial $M$ remains $<0.1 M_0$ in the whole range of magnetic fields accessible in the PNR system. In Fig.~\ref{f:f2}c we plot magnetization in the bulk of EuSe film and in the interfacial layer measured at $H=1$ T. The bulk magnetization gradually decreases with temperature for $T>T_N$, while the interfacial magnetization remains close to the noise level up to at least 25 K, $\times 5$ the N\'eel’s temperature of the bulk EuSe.

\begin{figure}[t]
\centering
\includegraphics[width=0.98\textwidth]{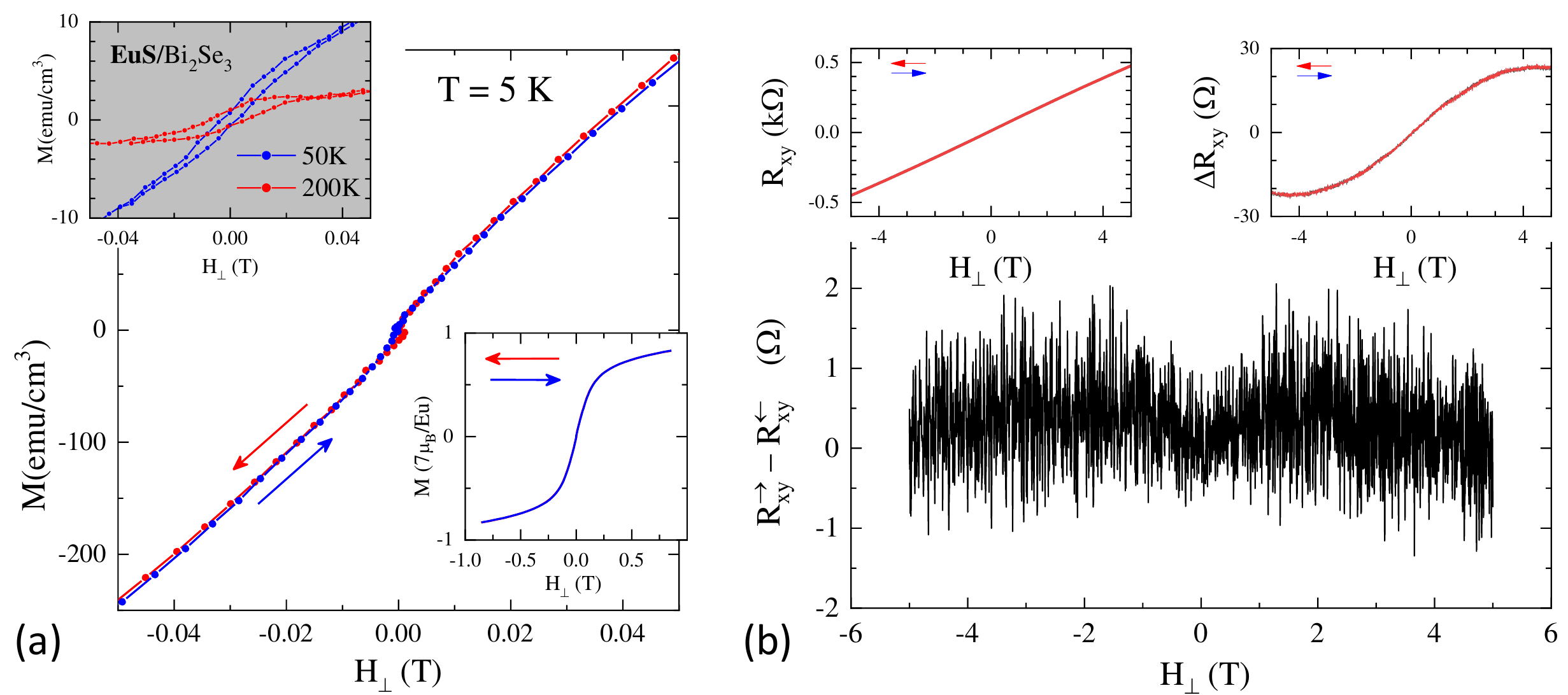}
\caption{\textbf{Out-of-plane magnetization studies.} Magnetization and Hall measurements are performed at 5 K, where the bulk of EuSe is in the paramagnetic phase. (a) Out-of-plane magnetization as a function of magnetic field is measured using SQUID magnetometer. The paramagnetic response of EuSe is clearly seen in the enlarged $H_\bot$ range in the lower right inset. The inset with a grey background reproduces the data for EuS/\BS\ films taken from Ref.~\onlinecite{Katmis2016} which shows pronounced hysteresis. (b) Difference between Hall resistances measured for two sweep directions shows no hysteresis. The Hall resistance ($R_{xy}$) and the Hall resistance with a linear component subtracted ($\Delta R_{xy}$) are plotted in the insets.}	
\label{f:f3}
\end{figure}

The sharpness of \BSES\ interface over the whole macroscopic sample, revealed by PNR, XRR and HRTEM, combined with 0.5 nm depth resolution of the PNR technique \cite{LauterPasyuk2007,Toperverg2016} excludes surface roughness and/or intermixing at the \BSES\  interface as a possible reason for the magnetization reduction near the interface. Alternative scenarios for spin arrangement with zero net in-plane magnetization include (i) formation of an out-of-plane ferromagnetic state, or (ii) formation of an antiferromagnetic state. The  thickness of the interfacial layer is $\approx 10\%$ of the total thickness of the EuSe film and in the case a ferromagnetic layer is formed near the interface we expect this layer to account for 10\% of the saturation magnetization of 1100 emu/cm$^3$ ($4.5\cdot10^{-4}$ emu for our sample), well within the $10^{-8}$ emu resolution of a SQUID magnetometer. Indeed, in EuS/\BS\  films hysteresis is clearly observed in the magnetization loops in the paramagnetic EuS phase up to 200 K; this hysteresis is considered as an unambiguous evidence that a ferromagnetic state is formed at the interface\cite{Katmis2016}. Magnetization loops of \BSES, measured at $T = 5$ K, are shown in Fig.~\ref{f:f3}a, where no hysteresis is observed within a resolution of a few emu/cm$^3$. For a ferromagnetic EuSe layer 1-2 nm thick we expect a loop with a magnetization amplitude of 50-100 emu/cm$^3$. The absence of hysteresis in magnetization loops in \ESBS\ is contrasted with a pronounced hysteresis in EuS/\BS\ reported in Ref.~\onlinecite{Katmis2016} and is consistent with prior studies of \BSES\cite{Prokes2021}. The Hall resistance also shows no hysteresis, see Fig.~\ref{f:f3}b, which indicates that no ferromagnetic state is formed at the interface. We observe anomalous Hall effect (AHE) (inset in Fig.~\ref{f:f3}b), which is most likely originates from the magnetization of the bulk EuSe. Similar AHE was observed in EuS/\BS\ samples\cite{Katmis2016}.

\begin{figure}[t]
\centering
\includegraphics[width=0.98\textwidth]{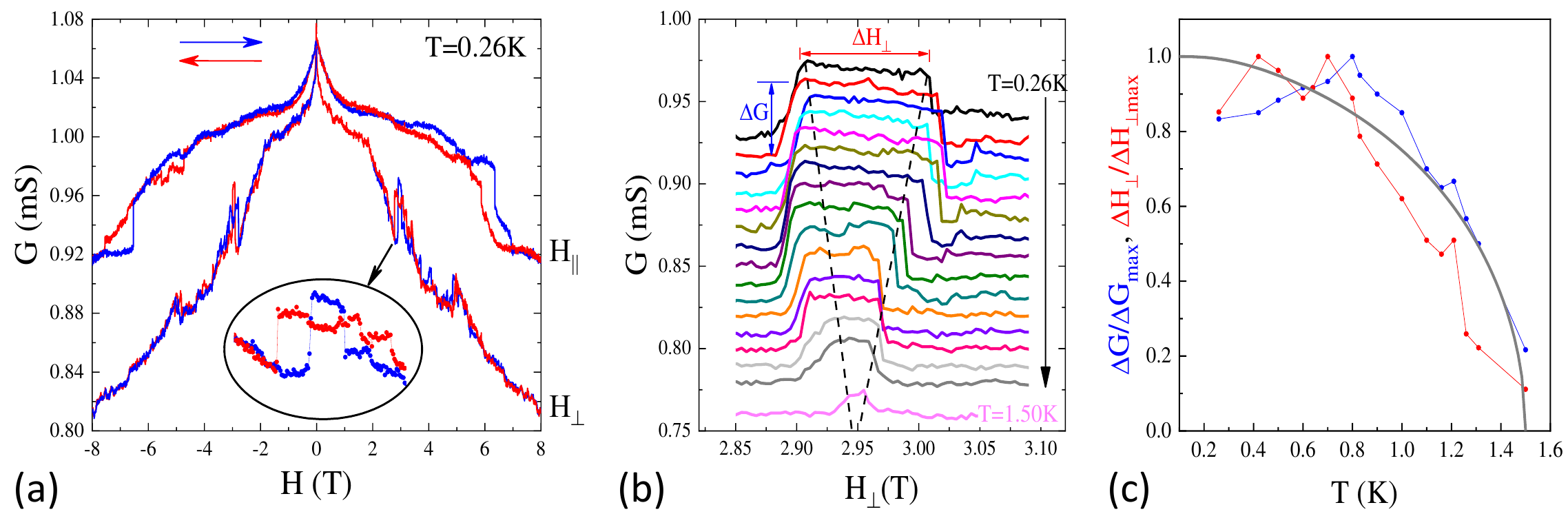}
\caption{\textbf{Sharp switching of sample conductance at high magnetic fields} (a) Two-terminal conductance $G$ of a microscopic sample 1$\mu m\times 2\mu m$ is plotted as a function for in-plane ($H_\|$, top curves) and out-of-plane ($H_\bot$, bottom curves) magnetic fields. In the inset a small region with sharp changes of $G$ is zoomed in. (b) Temperature dependence of the sharp switching near $H_\bot=2.95$ Tesla for the sweep-up direction. The curves are offset by -0.01 mS relative to the $T=0.26$ K curve. (c) Temperature dependence of the width and height of the switching region, normalized to their maximum values, is plotted as a function of temperature. The grey line is a mean field magnetization for $S=7/2$ and $T_c=1.5$ K.}	
\label{f:f4}
\end{figure}

The lack of magnetization response to the in-plane magnetic field measured by PNR and an absence of an out-of-plane ferromagnetic state points to the formation of an antiferromagnetic state at the interface. The transport data further corroborates the formation of a correlated magnetic state. In Fig.~\ref{f:f4}a we plot magnetoresistance measured in a mesoscopic device. Enhancement of conductivity at low magnetic fields is due to weak antilocalization, and smooth slow reduction of conductivity is expected at higher fields. Unexpectedly, several sharp features are observed at high fields $H>2$ T for both in-plane $H_\|$ and out-of-plane $H_\bot$ field directions. These fields are much larger than $\approx 0.2$ T needed to fully saturate magnetization in the bulk of the EuSe film. Most of the switchings are abrupt and hysteretic, as seen in the zoomed data. Qualitatively, the observed switchings are reminiscent of resistance jumps in mesoscopic magnetic materials where field-induced reconstruction of magnetic domains affects the current path.  The temperature dependence for one of such switching events, measured for the up-sweep of magnetic field, is analyzed in Fig.~\ref{f:f4}(b,c). Both the amplitude and the width of the switching approximately follow the temperature dependence of the mean field magnetization $M(T)/M(0)=B_s[3S/(S+1)\cdot T_c/T\cdot M(T)/M(0)]$, where $B_s$ is the Brillouin function. $M(T)/M(0)$ is plotted as a grey line in (c) where we used $S=7/2$ and $T_c=1.5$ K. This temperature scaling reiterates the magnetic origin of the switching events.

\begin{figure}[t]
\centering
\includegraphics[width=0.70\textwidth]{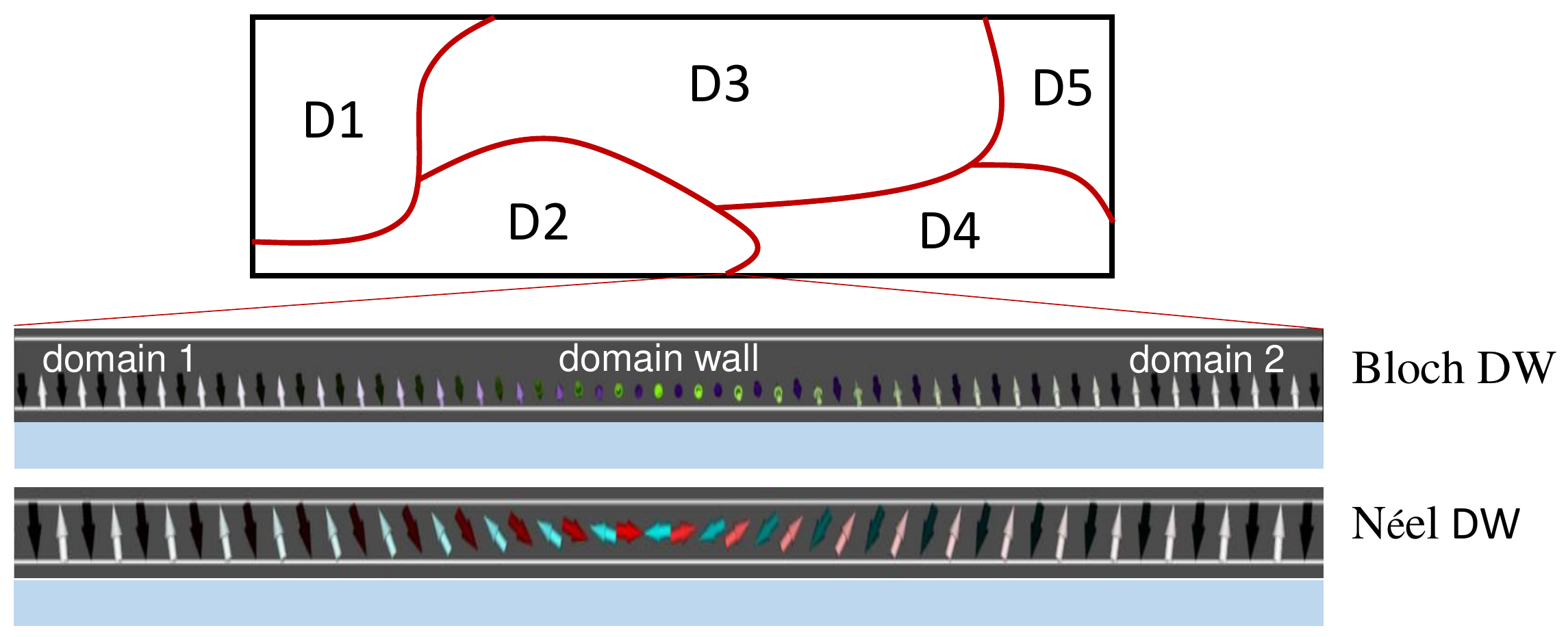}
\caption{\textbf{Antiferromagnetic domains and domain walls.} Top: schematic of a multi-domain sample. Bottom: micromagnetic simulations of extended Bloch and N\'eel domain walls at a boundary between two antiferromagnetic domains with out-of-plane magnetization.}	
\label{f:f5}	
\end{figure}

Epitaxial EuSe films are very resistive: when a vertical voltage $\pm0.1$ V is applied across the 20 nm film using mm-size contacts the leakage current does not exceed 0.5 nA. The EuSe film forms a multi-domain structure with the lateral domain size $\sim20$ nm, see an extended TEM image in the Supplementary material. At $H=0$ and $T<T_N$ EuSe is in an AF state and we anticipate formation of vertical magnetic domain walls (DW) between physical domains. The measured conductance $<5$ nS across the EuSe film includes transport through the bulk and along the DWs. Thus $40\ \mu$S jumps observed in magnetoresistance, Fig.~\ref{f:f4}, cannot originate from the transport through the EuSe layer. We conclude that magnetic reconstructions in the interfacial EuSe layer affect transport in the \BS\ TSS.

In Fig.~\ref{f:f5} we model spin arrangements in the vicinity of a Bloch or a N\'eel domain wall formed between two antiferromagnetic domains with out-of-plane magnetization axis. In both cases spins within domain walls rotate in-plane, yet neighboring spins remain nearly parallel.
Formation of multiple domains results in the formation of a network of domain walls, as shown schematically in Fig.~\ref{f:f5}.  Provided that TSS conductivity is different beneath domains and DWs, an abrupt reconstruction of domain structure leads to an abrupt reconstruction of conducting channels. Experimentally, such scenario is consistent with an abrupt magnetoresistance changes, as well as with PNR and magnetization data. Whether exchange interaction between an AF and a TI can affect transport in TSS or even open a gap in the TSS energy spectrum requires further theoretical investigation.

\section*{Methods}

\textbf{Material growth.} Multiple \BSES\ heterostructures are grown using molecular beam epitaxy (MBE), details of the growth, structural and magnetic characterization of EuSe films grown on different substrates can be found in Ref.~\onlinecite{Wang2022a}. For this study we use EuSe(001)/\BS (20nm/16QL) grown on a sapphire substrate, HRTEM images can be found in the Supplementary material.

\textbf{PNR \& XXR.} PNR experiments were performed on the Magnetism Reflectometer at the Spallation Neutron Source at Oak Ridge National Laboratory \cite{Lauter2009}, using neutrons with wavelengths $\lambda$ in the range of 0.25 – 0.85 nm and a high polarization of 98.5–99\% of the neutron beam. The measurements were carried out in a closed cycle refrigerator (Advanced Research System) equipped with a 1.15 T Bruker electromagnet and/or a 5 T cryomagnet. Using the time-of-flight method, a collimated polychromatic beam of polarized neutrons with a wavelength range  $\Delta\lambda$ is incident on the film at a grazing angle $\theta$, interacting with atomic nuclei and spins of unpaired electrons. The reflected intensities $R^+$ and $R^-$ are measured as a function of the wave vector transfer, $Q = 4\pi\sin(\theta)/\lambda$, with the neutron spin parallel (+) or antiparallel (-), respectively, to the applied field. To separate nuclear scattering from magnetic scattering, the spin asymmetry ratio $SA = (R^+ – R^-)/(R^+ + R^-)$ is calculated, for which $SA = 0$ means the absence of a magnetic moment in the system. Being electrically neutral, spin-polarized neutrons penetrate the entire multilayer structure and probe the magnetic and structural composition of the film and hidden interfaces down to the substrate. PNR is a deep penetrating depth sensitive technique that examines the chemical and magnetic depth profiles of materials with a resolution of 0.5 nm. The depth profiles of the nuclear and magnetic scattering length densities (NSLD and MSLD) correspond to the depth profile of the chemical and in-plane magnetization vector distributions, respectively.

XRR measurements were performed at the Center for Nanophase Materials Sciences (CNMS), Oak Ridge National Laboratory. XRR measurements were conducted on a PANalytical X’Pert Pro MRD equipped with hybrid monochromator and Xe proportional counter. For the XRR measurements, the X-ray beam was generated at 45 kV/40 mA, and the X-ray beam wavelength after the hybrid mirror was $\lambda=1.5406$ \AA (Cu K$\alpha$1 radiation).

\textbf{Transport measurements.} For transport studies devices in Hall bar geometry have been fabricated using conventional e-beam lithography and ion milling. Ohmic contacts on mesoscopic samples were formed by ion milling the top EuSe layer and deposition of Ti/Au or Nb/NbN films. Transport measurements were performed in a variable temperature $^3$He refrigerator with 8 T magnet using standard low frequency lock-in techniques.

\section*{Acknowledgments}

Authors acknowledge support by the NSF DMR-2005092 (Y.W., L.P.R), and NSF DMR-1905277 (X.L., B.A.A.) grants. M.Z. and T.O. acknowledge use of facilities for High Resolution Electron Microscopy at University of Notre Dame. This material is based upon work supported by the U.S. Department of Energy, Office of Science, National Quantum Information Science Research Centers, Quantum Science Center. This research used resources at the Spallation Neutron Source, a Department of Energy Office of Science User Facility operated by the Oak Ridge National Laboratory. XRR measurements were conducted at the Center for Nanophase Materials Sciences (CNMS), which is a DOE Office of Science User Facility.

The data that support the findings of this study are available from the corresponding author upon reasonable request.

\section*{Author contributions}

L.P.R., J.K.F. and X.L. conceived the experiments, Y.W. performed magnetization and transport measurements, V.L. performed PNR measurements and analysed PNR and XRR data, H.A took part in PNR experiments, J.K carried XRR measurements, J.W. performed high resolution XRD measurements, S.T.K. and P.U. performed micromagnetic simulations, M.Z and T.A.O performed HRTEM. Y.W., V.L., O.M., S.T.K, B.A.A, X.L. and L.P.R. written the manuscript.

\bibliography{rohi}


\clearpage
\newpage
\onecolumngrid

\renewcommand{\thefigure}{S\arabic{figure}}
\renewcommand{\theequation}{S\arabic{equation}}
\renewcommand{\thetable}{S\arabic{table}}
\renewcommand{\thepage}{sup-\arabic{page}}
\setcounter{page}{1}
\setcounter{equation}{0}
\setcounter{figure}{0}
\setcounter{table}{0}

\begin{center}
\textbf{\Large Discovery of a high-temperature antiferromagnetic state and transport signatures of exchange interactions in a \BSES\ heterostructure.}\\
\vspace{0.5cm}
\textbf{\large Supplementary Materials} \\
{\it Ying Wang, Valeria Lauter, Olga Maximova, Shiva T. Konakanchi, Pramey Upadhyaya, Jong Keum, Haile Ambaye, Jiashu Wang, Maksym Zhukovskyi, Tatyana A. Orlova, Badih A. Assaf, Xinyu Liu, and Leonid P. Rokhinson}
\end{center}

\tableofcontents

\clearpage
\section{Additional information on polarized neutron reflectometry measurements}

Unlike local probe techniques such as HTREM, atomic force microscope (AFM) or a scanning tunneling microscope (STM), PNR is a statistical method that probes the entire macroscopic lateral size of a sample with outstanding depth resolution of 0.5 nm \cite{LauterPasyuk2007,Toperverg2016} and sensitivity not only to the absolute value of magnetic moment \cite{Blundell_1995,Pasyuk1995}, but to the in-plane magnetization vector distribution \cite{Toperverg2001,Toperverg2005} and the domain structure \cite{LauterPasyuk2002,Lauter2003} through the depth of complex heterostructures \cite{Liu2011a}. Polarization analysis of the reflected and scattered beam can be performed using {\it in situ} polarized $^3$He system \cite{Tong2012,Jiang2013,Jiang2017} and a supermirror analyzer \cite{Syromyatnikov2014}.

\begin{figure}[h]
\centering
\includegraphics[width=0.98\textwidth]{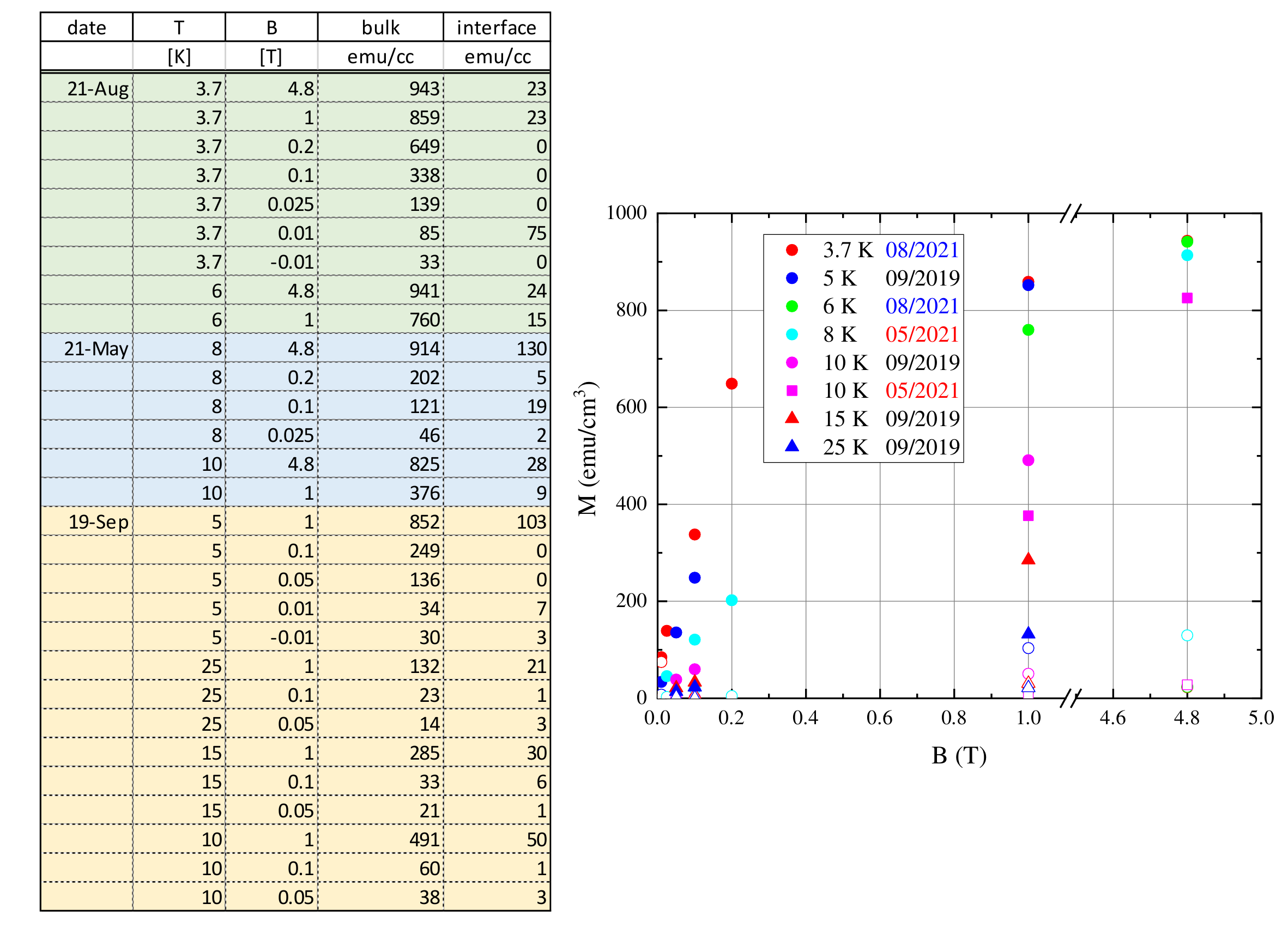}
\caption{PNR data was collected during three runs in Sept. 2019 and May and August 2021. The table lists bulk and interfacial magnetizations extracted from PNR data fittings. A summary plot includes bulk (filled symbols) and interfacial (open symbols) magnetization for all PNR scans.}	
\label{f:sf1}	
\end{figure}

The PNR data was collected during three runs over two years, parameters of the scans and extracted bulk and interface magnetizations are listed in Fig.~\ref{f:sf1}. During the 2019 run the lowest temperature and the highest field were limited to 5 K and 1 T respectively, the system was upgraded to allow temperatures down to 3 K and fields up to 5 T between the first and the second runs. During the second run a thermometer was not functioning properly and the sample temperature was determined by fitting bulk magnetization obtained by PNR to the $M(T,H)$ data obtained by SQUID magnetometry.

\clearpage
\section{Magnetization data and magnetic phase diagram of \BSES (TI/MI) films}

\begin{figure}[h]
\centering
\includegraphics[width=0.98\textwidth]{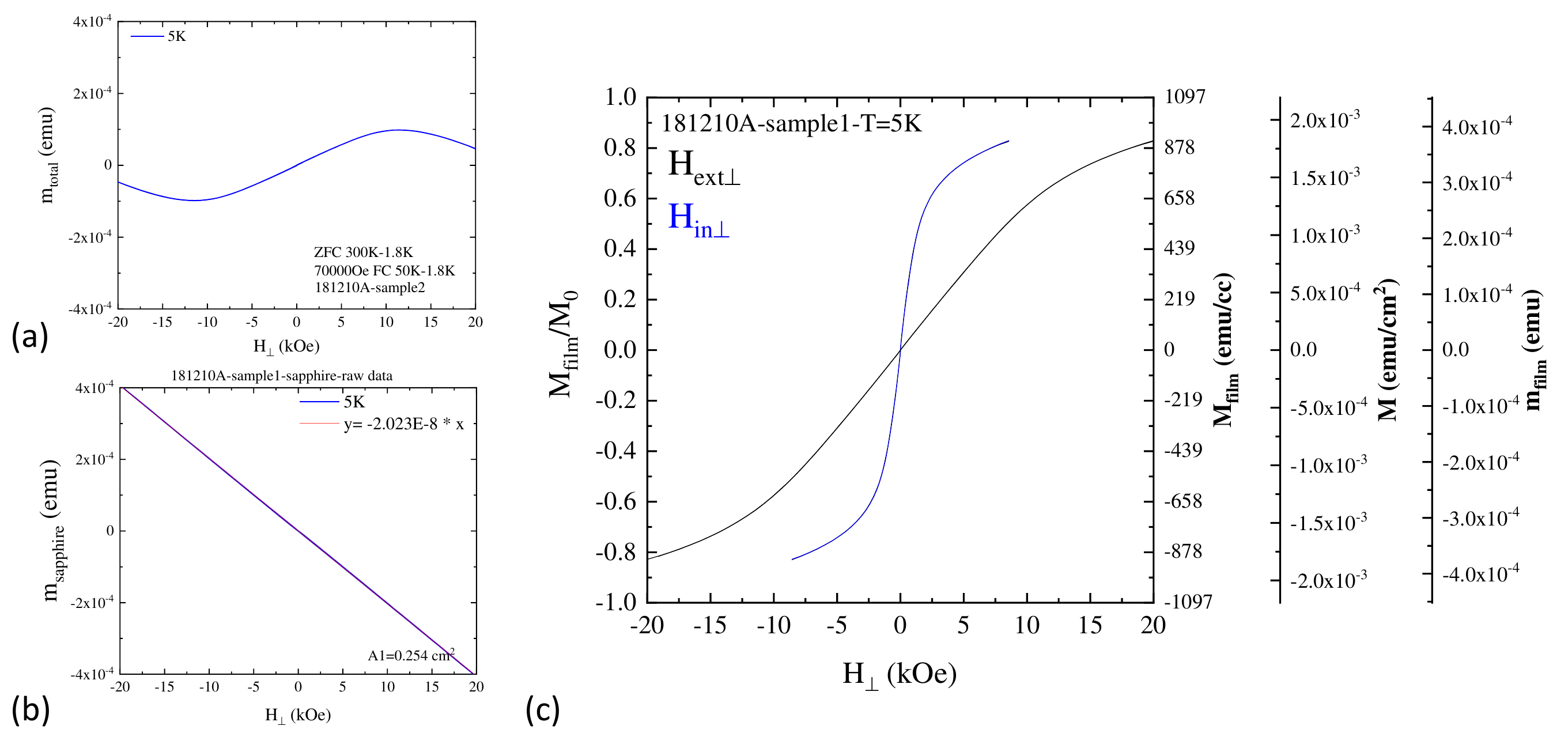}
\caption{Magnetization data processing. (a) Original data measured by a SQUID magnetometer at $T=5$ K. (b) Magnetization of a sapphire substrate measured at the same temperature. (c) Magnetization of the EuSe film is plotted as a function of external $H_\bot$ and internal $H_{int\bot}$ (corrected for a demagnetization factor $N=1$). $M_0=7\mu_B$ per Eu atom, where $\mu_B$ is a Bohr magneton.}	
\label{f:sf2}	
\end{figure}

The magnetic properties of bi-layer hetero-structure \ESBS\  were measured by a Quantum Design MPMS-3 SQUID magnetometer. Samples were carefully handled using plastic tweezers and cleaned with isopropanol followed by N2 blow dry. We used
standard plastic translucent drinking straws as wafer holders and mounted samples in the straws such that the magnetic field was applied either in-plane ($H_{\parallel}$) or out-of-plane ($H_{\perp}$).

Due to remnant
magnetic field in the superconducting magnet (See Quantum Design Application Note 1500-011), a correction to the reported applied magnetic field was applied to all the magnetization loops data. This field correction was measured and verified using a single crystal of gadolinium gallium garnet, which is a clean paramagnetic sample and known to have a linear and reversible magnetization curve. For the field sweep rate used in this paper the real field is found to be lagging by 25 Oe. Furthermore, by considering the demagnetization effect, all magnetization measurements were corrected to account for the demagnetizing
field of thin film (demagnetization factors $N_{\parallel}=0$ and $N_{\perp}$=1).

In Fig.~\ref{f:sf2}(a,b) we display unprocessed magnetization data for \ESBS/sapphire and sapphire at $T=5$ K. Magnetization of the sapphire substrate was measured after removal of the \ESBS\ film with a plastic blade. Magnetization of epilayers \ESBS\ is
obtained by subtracting the substrate magnetization $M_{film} = m_{total} - m_{sapphire}$. In Fig.~\ref{f:sf2}(c) magnetization loops for out-of-plane direction of magnetic field are plotted in different units (total magnetization emu, areal magnetization emu/cm$^2$, volume magnetization emu/cm$^3$ and atomic magnetization per Eu atom $M/M_0$). To convert volume to atomic magnetization we used the lattice spacing $\lambda_{EuSe}=6.186 \AA$ measured by XRD and expected magnetization of fully polarized $Eu^{2+}$ ion $M_0=6.94 \frac{\mu_B}{Eu}$.

\begin{figure}[h]
\centering
\includegraphics[width=0.99\textwidth]{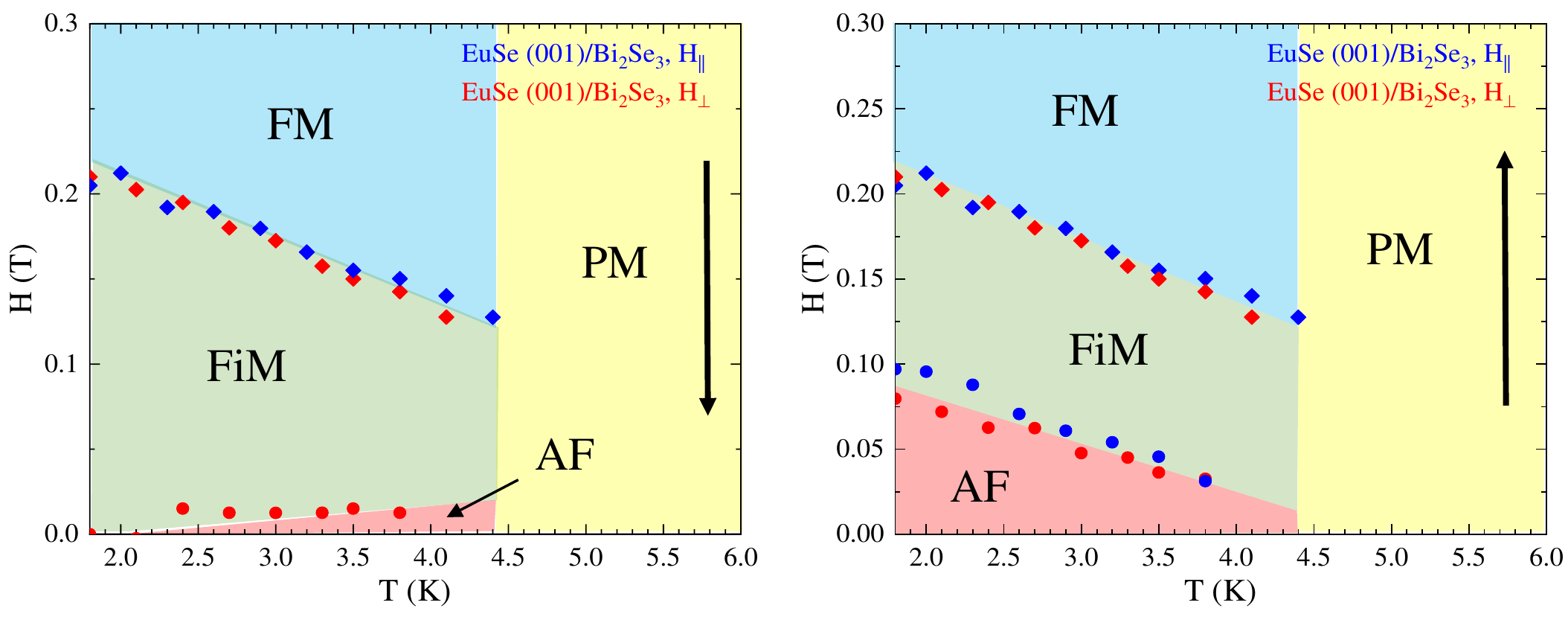}
\caption{Phase diagram of a thin \ESBS\  film studied in the main text for magnetic field scans from high to low fields. The transition points (dots on the plot) are extracted from magnetization curves measured with a SQUID magnetometer. Black arrows indicate field sweep direction.}	
\label{f:sf3}	
\end{figure}

Several magnetic phase transitions are observed below $T_N=4.4$ K, these transitions have been studied in our previous publication \cite{Wang2022a}. In Fig.~\ref{f:sf3} we reproduce a phase diagram for the sample discussed in the main text. Note that AF-FiM transition is hysteretic, while position of the FiM-FM does not depend on the field sweep direction and bulk of EuSe is fully polarized for $H>0.2$ T.

\clearpage
\section{High resolution TEM images}

\begin{figure}[h]
\centering
\includegraphics[width=0.9\textwidth]{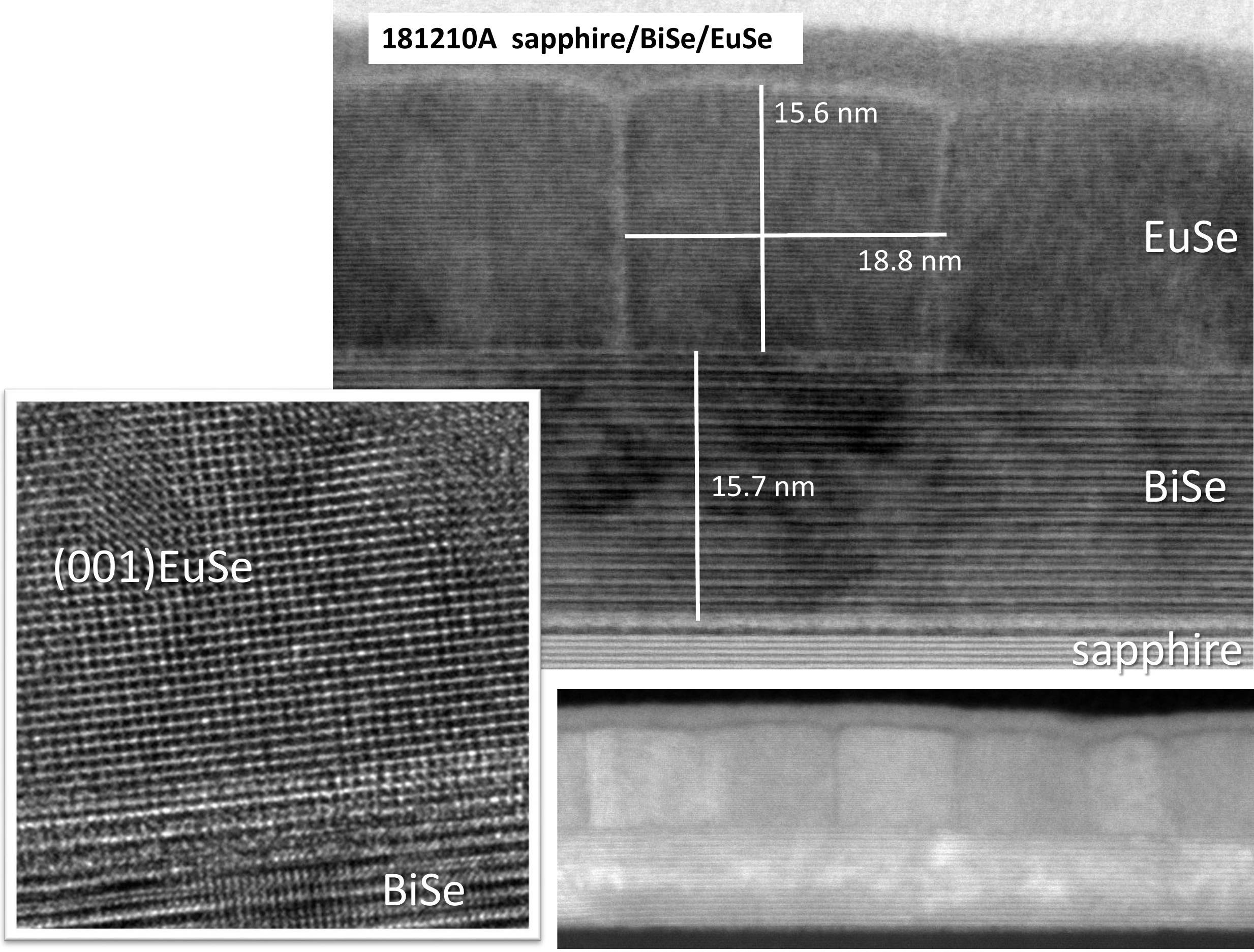}
\caption{High resolution TEM images of \BSES\ heterostructure grown on a sapphire substrate. \BS\ growth in (0001) orientation and \BSES\ interface roughness is $<1$QL of \BS. EuSe growth primarily in (001) direction, as can be clearly seen in HRTEM images and confirmed by high resolution XRD studies\cite{Wang2022a}. EuSe forms domains with size 15-30 nm as can be seen in a low resolution TEM image.}	
\label{f:sf4}	
\end{figure}

\clearpage
\section{Transport properties of the EuSe layer}

\begin{figure}[h]
\centering
\includegraphics[width=0.9\textwidth]{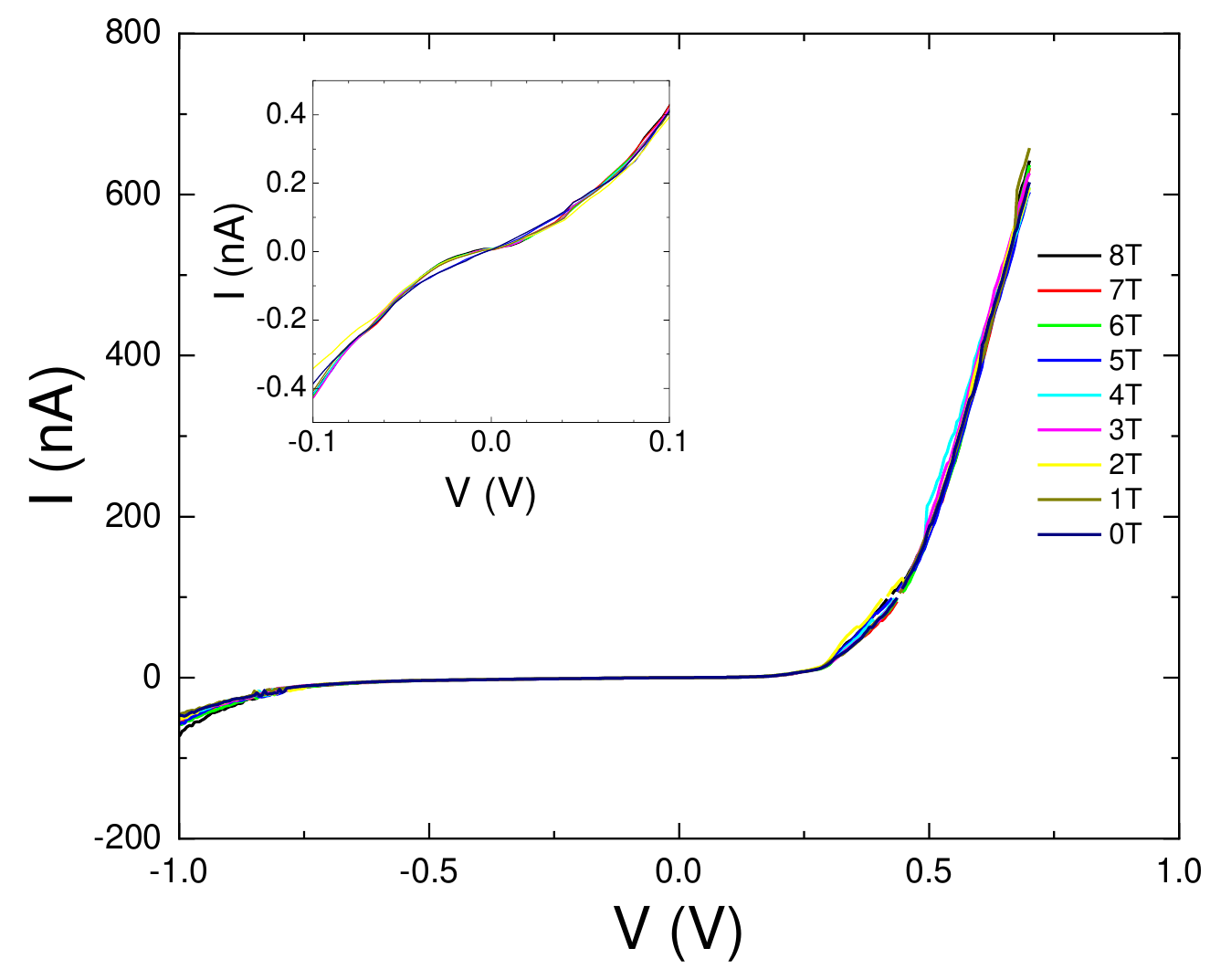}
\caption{Vertical transport across EuSe layer. \BS\ film acts as a bottom contact, at the top of the EuSe film there is a large (few mm$^2$) silver paste contact.}	
\label{f:sf6}	
\end{figure}

\clearpage
\section{Details of Micromagnetic Simulations}

We used a grid size of $128$x$1$x$1$ with a cell size of $1$ nm in the micromagnetics software mumax \cite{MuMax2014} to simulate $1$D domain walls in $EuSe$. Saturation magnetization $M_s = 1100$ emu/cm$^3$, exchange stiffness $A=-50$ pJ/m and perpendicular anisotropy $K_u=0.38$ MJ/m$^3$ were used as model parameters. In the absence of a domain wall, the spins equilibriate into an antiferromagnetic configuration with each spin pointing (either up or down) in the direction opposite to its immediate neighbors. To simulate Bloch domain wall, we reorient the two spins at the center of the chain to point into the page and out of the page directions respectively. We then let mumax relax the spin chain to find its lowest energy configuration resulting in the Bloch domain wall shown in Fig.~\ref{f:f5}. To simulate Neel domain wall, we reorient the two spins at the center of the spin chain in equilibrium to point along left and right directions respectively and let the chain relax into its lowest energy configuration. In principle both Bloch and Neel domain walls will have in-plane spins at the center of the domain wall.

\end{document}